\documentclass[prb,reprint,aps,twocolumn]{revtex4-1}

\pdfoutput=1

\usepackage[utf8]{inputenc}

\usepackage[T1]{fontenc}
\usepackage{hyperref}
\usepackage{color}
\usepackage{graphicx}
\usepackage{units}
\usepackage{amsmath}
\usepackage{float}

\usepackage{epstopdf}
\usepackage{subcaption}
\usepackage{amssymb}
\usepackage{pifont}
\usepackage{textcomp}
\usepackage[usenames,dvipsnames]{xcolor}
\usepackage{ulem}
\usepackage{amsmath}

\hypersetup{colorlinks = true, citecolor = blue, breaklinks = true}
\usepackage[usenames,dvipsnames]{xcolor}
\usepackage{gensymb}
\usepackage{xspace}
\begin{document}

\title{Band-gap engineering in AB(O\textsubscript{x}S\textsubscript{1-x})\textsubscript{3}  perovskite oxysulfides: A route to strongly polar materials for photocatalytic water splitting}
\date{\today}
\author{Nathalie Vonr\"uti}
\affiliation{Department of Chemistry and Biochemistry, University of Bern, Freiestrasse 3, CH-3012 Bern, Switzerland}
\author{Ulrich Aschauer}
\affiliation{Department of Chemistry and Biochemistry, University of Bern, Freiestrasse 3, CH-3012 Bern, Switzerland}

\begin{abstract}
Polar heterogeneous photocatalysts were shown to lead to enhanced charge-carrier separation that results in superior activity for example for photocatalytic water splitting. Promising photocatalyst materials such as oxynitrides can be rendered polar by epitaxial strain, which however also increases their band gap, making them unsuitable for visible light absorption. This suggests a trade-off between small band gaps and polar distortions - both being crucial for the catalyst's efficiency. In this paper we investigate, using density functional theory calculations, the suitability of strained AB(O\textsubscript{x}S\textsubscript{1-x})\textsubscript{3} perovskites for photocatalytic water splitting. These materials normally have band gaps too small for water splitting but inducing polar distortions via epitaxial strain can increase the band gap to the suitable range. We find perovskite BaZr\textsubscript{y}Ti\textsubscript{1-y}O\textsubscript{2}S compounds to be highly promising for photocatalytic water splitting due to large polar distortions and suitable band gaps.
\end{abstract}

\maketitle

\section{Introduction}

Efficient electron-hole separation is a key requirement for high-performing photocatalysts for example for photocatalytic water splitting. Charge-carrier separation can be enhanced by polar materials such as ferroelectrics, where the compensation of the internal dipolar field results in preferential migrations of electrons and holes in opposite directions \cite{giocondi2001spatial, li2014photocatalysts}. Ferroelectric materials, which traditionally are d\textsuperscript{0} perovskite oxides (ABO\textsubscript{3}), \cite{matthias1949new, hill2000there} typically have large band gaps above 3 eV. Optimal band gaps for photocatalytic water splitting, however, are much lower: While very small band gaps would lead to absorption of a large fraction of the solar spectrum, the condition that the valence and conduction band edges have to straddle the oxygen and hydrogen evolution potential respectively, imposes a minimal band-gap of 1.23 eV. After considering theoretical minimum overpotentials for the oxygen and hydrogen evolution reaction, this minimum increases to 1.7 eV and since in practice there are always energy losses such as electrode polarization, band gaps should be larger than approximately 2 eV \cite{bolton1985limiting}. One possibility to reduce the large band gap of oxides is by substitution of oxygen with nitrogen, which has a lower electronegativity ($\eta=3.04$) than oxygen ($\eta=3.44$) resulting in materials with higher valence band edges and band gaps just above 2 eV \cite{kasahara2002photoreactions}. 
 
\begin{figure}
	\includegraphics[width=0.9\columnwidth]{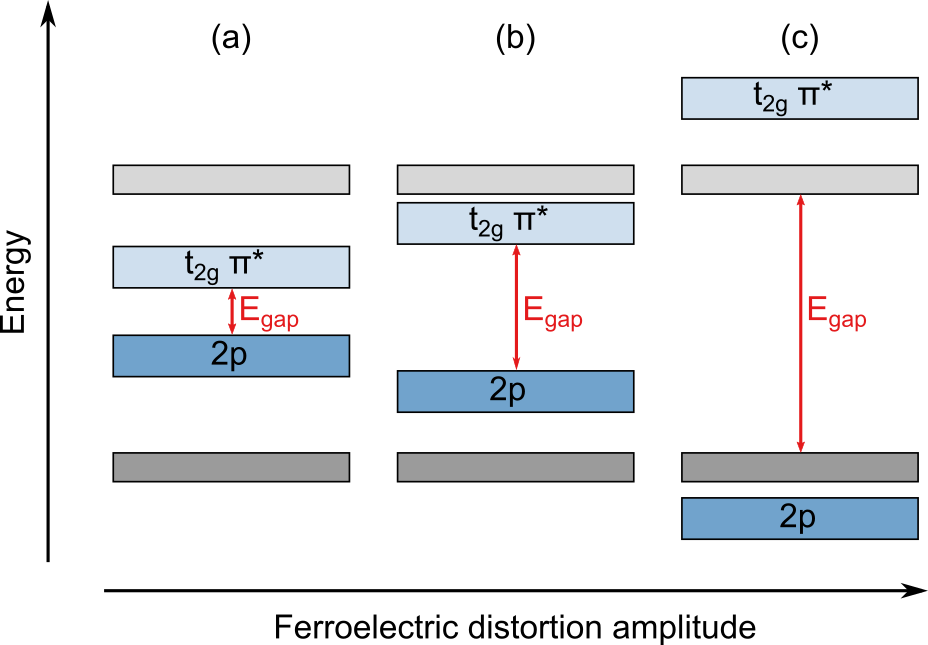}
	\caption{Schematic representation of changes in band energies and resulting band gaps for increasing ferroelectric distortion amplitudes. The band gap increases from (a) small to (b) intermediate distortion amplitudes and (c) remains constant above a certain amplitude. Blue and grey boxes represent bands that are affected (2p and t\textsubscript{2g} $\pi$*) respectively not affected by the ferroelectric distortion and the darker/lighter shade distinguishes occupied/empty states respectively.}
	\label{fig:sketch_bandgap}
\end{figure}

Oxynitrides were not reported as ferroelectric in their ground state but out-of-plane ferroelectricity can be induced in this class of material using compressive biaxial strain, \cite{oka2014possible, vonruti2018anion} applied for example by lattice matching during coherent epitaxial growth by pulsed layer deposition or molecular beam epitaxy. Whereas the in-plane lattice parameters shorten with increasing compressive strain, the out-of-plane lattice parameter expands due to Poisson's effect. This elongation affects the balance between coulombic and covalent interactions and eventually leads to ferroelectric distortions along the elongated direction \cite{bersuker2013pseudo, rondinelli2009non}. Ferroelectric distortions in oxides and oxynitrides however also increase the band gap due to a stabilizing effect on the O 2p respectively N 2p orbitals forming the valence-band edge together with a destabilization of the transition-metal t$_{2g}$ $\pi^*$ orbitals forming the conduction-band edge \cite{wheeler1986symmetric,vonruti2018epitaxial}. The band gap in strained oxide and oxynitride materials hence depends on the strain-dependent existence and amplitude of the ferroelectric distortion as illustrated in Fig. \ref{fig:sketch_bandgap} (a) and (b). While this effect is rather small in oxides \cite{berger2011band}, we have shown it to be much more pronounced in oxynitrides, where the reduced electronegativity of nitrogen leads to stronger bonds and hence larger ferroelectric distortions that increase the band gap by up to 1.5 eV at 4\% compressive strain, effectively disabling visible light absorption \cite{vonruti2018epitaxial}.

In search for polar materials with suitable band gaps, we investigate in this study perovskite (oxy)sulfides. Due to band gaps below 2 eV resulting from the low electronegativity of sulfur ($\eta=2.58$), this class of materials has so far mainly been investigated for photovoltaics \cite{sun2015chalcogenide, ju2017perovskite} or for Z-scheme water splitting \cite{sun2018efficient, ma2016visible}. A polar distortion, induced for example by epitaxial strain, could however increase the band gap to slightly above 2 eV and yield materials suitable for photocatalytic water splitting.

Besides the smaller electronegativity, sulfur also has a much larger ionic radius \cite{shannon1976revised} (r=1.84 \AA) compared to oxygen (r=1.40 \AA) and nitrogen (r=1.46 \AA). Considering the tolerance factor and Pauling's first rule, this leads to a limited number of possible elements on the A and B site in the perovskite ABS\textsubscript{3} structure. Indeed only compounds with A = Ca, Sr, Ba, Eu and B = Zr, Hf, U  have been reported in the GdFeO\textsubscript{3} perovskite structure \cite{lelieveld1980sulphides}. Mixed-anion materials with both sulfur and oxygen were reported to be difficult to synthesize and mainly crystallize in Ruddlesden-Popper structures where apical and equatorial anion sites with different coordination environments exist\cite{kobayashi2018property}.

We therefore start our density functional theory (DFT) investigation of the AB(O\textsubscript{x}S\textsubscript{1-x})\textsubscript{3} system with BaZrS\textsubscript{3}, which is one of most investigated perovskite sulfides \cite{perera2016chalcogenide, niu2017bandgap, bennett2009effect, meng2016alloying, brehm2014structural} and for which sulfur was partially substituted with oxygen \cite{perera2016chalcogenide}. We then systematically investigate polarity and band gaps in related compounds obtained by partial substitution of sulfur by oxygen as well as by changing the A site (Ca, Sr, Ba) and the B site (Ti, Zr, Hf). We find the BaZr\textsubscript{y}Ti\textsubscript{1-y}O\textsubscript{2}S system to be most promising in terms of small band gaps ($E_g = 2.8\:eV$ at the HSE level of theory for BaZr\textsubscript{0.625}Ti\textsubscript{0.375}O\textsubscript{2}S under 3\% compressive strain). The actual band gap could even be slightly lower as experiments for BaZrO\textsubscript{2}S \cite{perera2016chalcogenide} indicate that the measured band gap is systematically lower than HSE predictions, which could be related to a small degree of anion disorder. These materials also have a strong out-of-plane polarization but are unlikely to be ferroelectric due to their large theoretical coercive fields.


\section{Computational methods}

We performed DFT calculations with Quantum ESPRESSO \cite{giannozzi2009quantum} at the PBE+\textit{U} level of theory \cite{perdew1996generalized, anisimov1991band}. Hubbard \textit{U} was calculated self-consistently \cite{hsu2009first, timrov2018hubbard} for the different transition metal B sites resulting in 3.1 eV for Ti, 1.8 eV for Zr and 1.5 eV for Hf. While PBE+\textit{U} band gaps severely underestimate experimental band gaps, they were shown to follow the same relative strain-dependence as quasiparticle GW band gaps \cite{berger2011band}. In order to compare band gaps between different compositions, we selectively performed hybrid functional calculations using HSE ($\alpha = 0.18$) on structures relaxed at the PBE+\textit{U} level.

Calculations of the perovskite structures were performed in a 40 atom $2\times 2\times 2$ supercell of the 5-atom unit cell. Reciprocal space was sampled with a $4\times 4\times 4$ Monkhorst-Pack grid \cite{monkhorst1976special}. All atoms are represented by ultrasoft pseudopotentials \cite{vanderbilt1990soft} with Ca(3s, 3p, 4s, 3d), Sr(4s, 4p, 5s), Ba(5s, 5p, 5d, 6s, 6p), Ti(3s, 3p, 3d, 4s), Zr(4s, 4p, 5s, 5p, 4d) Hf(5s, 6s, 5p, 6p, 5d), O(2s, 2p) and S(3s, 3p, 3d) valence electrons. The cutoff for the plane-wave basis set was 40 Ry for the kinetic energy combined with 400 Ry for the augmented density. The convergence criteria for geometry relaxations were 0.05 eV/\AA\ for forces and $1.4\cdot 10^{-5}$ eV for the total energy. 

Epitaxial strain imposed by the (001) facet of a cubic substrate is modelled by setting the in-plane lattice parameters to equal lengths and orthogonal to each other. Strain is defined with respect to the average in-plane lattice constant of the unstrained structure. We used climbing-image nudged elastic band (cNEB) calculations \cite{henkelman2000climbing} to obtain the energy profile along the ferroelectric switching pathway. Paths in cNEB calculations were relaxed until the norm of the force orthogonal to the path was smaller than 0.05 eV/\AA\ for all images. The electric polarization was calculated using the Berry phase method \cite{king1993theory, resta1993macroscopic, vanderbilt1993d} with 6 k-points along the symmetry-reduced string. Crystal orbital Hamilton population (COHP) analysis \cite{dronskowski1993crystal, deringer2011crystal} was performed with LOBSTER \cite{maintz2013analytic, maintz2016efficient, maintz2016lobster} and atomic structures were visualized using VESTA \cite{momma2011vesta}.

\section{Results and discussion}

\subsection{BaZr(O\textsubscript{x}S\textsubscript{1-x})\textsubscript{3}}

We start our exploration of the AB(O\textsubscript{x}S\textsubscript{1-x})\textsubscript{3} system with the perovskite BaZrS\textsubscript{3}, which was reported to be antiferroelectric and to show large octahedral rotations \cite{bennett2009effect}. In oxides, a good descriptor for structural distortions such as octahedral rotations or ferroelectric distortion is the tolerance factor \cite{goldschmidt1926gesetze}. It is defined as
\begin{equation}
	t=\frac{r_A + r_X}{\sqrt{2}(r_B + r_X)}
	\label{eq:tolerance}
\end{equation}
where $r_A$ and $r_B$ are the ionic radii of the A and B site and $r_X$ is the radius of the anion. A tolerance factor $t>1$ implies an instability of the perovskite structure compared to hexagonal phases, a $t$ close to one suggests ferroelectricity, while for $t<1$ octahedral rotations become increasingly more favorable up to the point where the perovskite structure is no longer stable. For BaZrS\textsubscript{3} ($r_\mathrm{Ba}$=1.61 \AA, $r_\mathrm{Zr}$=0.72 \AA, $r_\mathrm{S}$=1.84 \AA)\cite{shannon1976revised}, the large radius of sulfur results in $t=0.95$, indeed suggesting large octahedral rotations.

\begin{figure}
	\includegraphics[width=0.9\columnwidth]{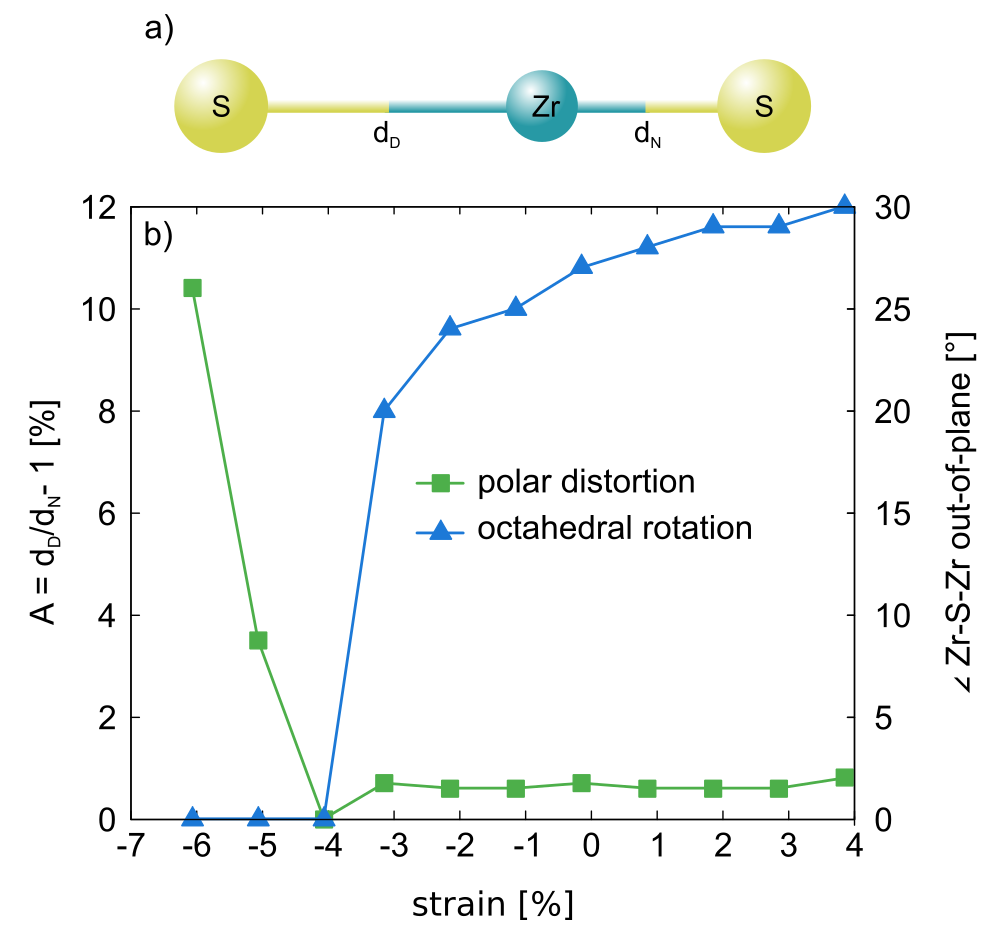}
	\caption{a) Definition of bond lengths used to describe the ferroelectric distortion and b) amplitude of the polar distortion and out-of-plane octahedral rotations in BaZrS\textsubscript{3} as a function of epitaxial strain.}
	\label{fig:ferro}
\end{figure}

We induce ferroelectricity in BaZrS\textsubscript{3} by applying
epitaxial strain and characterise the octahedral rotations and the polar distortion as these two distortions were shown to often compete with each other \cite{benedek2013there, aschauer2014competition}. We characterise the former by the deviation of the out-of-plane Zr-S-Zr bond angle from 180\textdegree\space, while the latter is measured via the deviation from 1 of the ratio of two consecutive S-Zr bonds along a given direction
\begin{equation}
	A = d_D/d_N - 1
\end{equation}
where $d_D$ and $d_N$ are the bond lengths of the longer and shorter S-Zr bond respectively as shown in Fig. \ref{fig:ferro}a). As shown in Fig. \ref{fig:ferro}b), we observe a gradual decrease of the out-of-plane octahedral rotations with increasing compressive strain. The polar distortion only appears once the octahedral rotations have vanished at a compressive strain between 4\% and 5\%. Such large strains are unlikely to be experimentally accessible. However, given the usual competition between polar distortions and octahedral rotations, it is likely that a reduction of the latter, for example by anionic or cationic alloying, will promote polarity at smaller compressive strain.

A partial substitution of sulfur by oxygen will increase the tolerance factor, which we calculate in a mixed-anion compound by using the arithmetic average of the two anion radii
\begin{equation}
r_X:=x\cdot r_O + (1-x)\cdot r_S
\end{equation}
with $r_\mathrm{O}$ = 1.40 \AA\ the ionic radius of oxygen\cite{shannon1976revised}. For the BaZr(O\textsubscript{x}S\textsubscript{1-x})\textsubscript{3} series we obtain  tolerance factors of 0.97, 0.99 and 1.00 for BaZrOS\textsubscript{2}, BaZrO\textsubscript{2}S and BaZrO\textsubscript{3} respectively, showing indeed an increase of $t$ and hence a reduction of the octahedral rotations with increasing oxygen content. 

\begin{figure}
	\includegraphics[width=0.9\columnwidth]{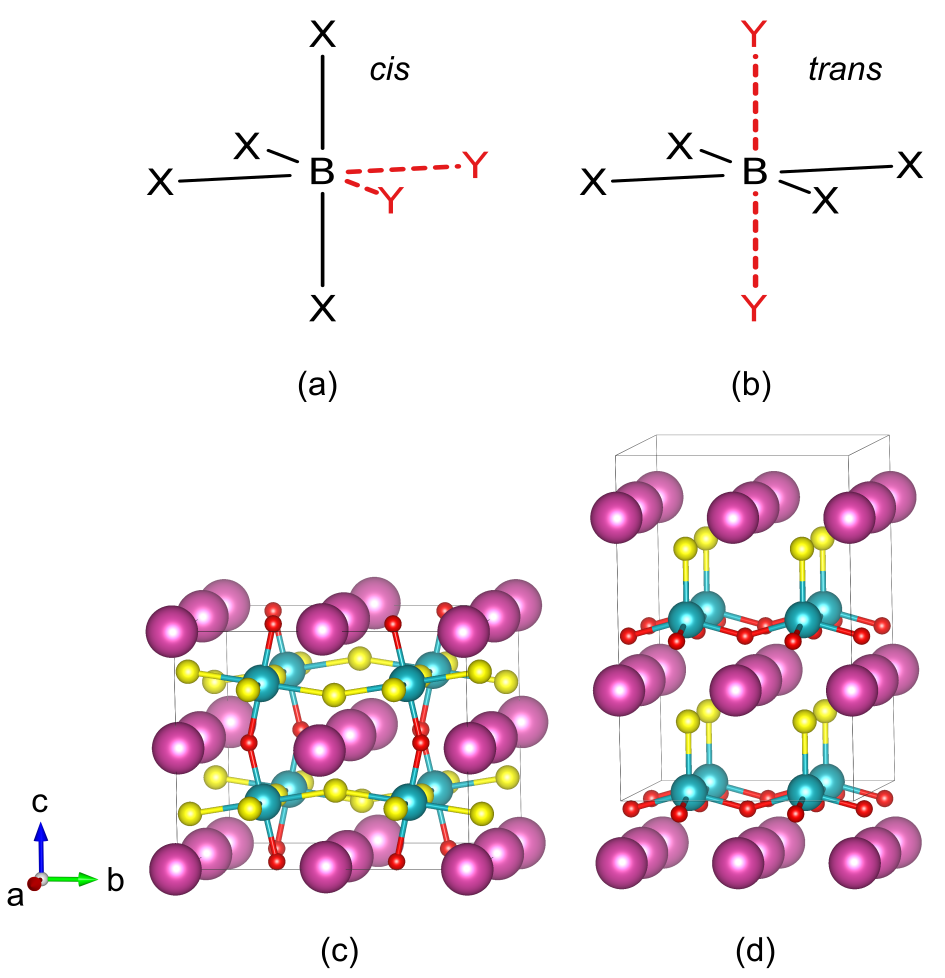}
	\caption{Schematic representation of the anion orders in a BX\textsubscript{4}Y\textsubscript{2} mixed-anion octahedron (a) \textit{cis} and (b) \textit{trans}. Relaxed structures of unstrained (c) BaZrOS\textsubscript{2} and (d) BaZrO\textsubscript{2}S. Color code: Ba=purple, Zr=cyan,  O = red, S= yellow }
	\label{fig:unstrained}
\end{figure}

In the mixed-anion compounds BaZrOS\textsubscript{2} and BaZrO\textsubscript{2}S, the oxygen and sulfur anions within one BX\textsubscript{6} octahedron can order either in \textit{cis} or \textit{trans} configuration shown in Fig. \ref{fig:unstrained}(a) and (b). We refer here to the anion configuration with respect to the minority anion species (O in BaZrOS\textsubscript{2} and S in BaZrO\textsubscript{2}S). While the \textit{cis} configuration is preferred in many mixed-anion perovskites such as oxynitrides \cite{attfield2013principles}, the \textit{trans} configuration is generally rare \cite{kobayashi2018property}, one example being SrVO\textsubscript{2}H, for which the hydride ions fully order on apical sites \cite{denis2014strontium}.

For both BaZrOS\textsubscript{2} and BaZrO\textsubscript{2}S we find the \textit{trans} order to be energetically much more favorable than the \textit{cis} order (0.23 eV/f.u. and 0.33 eV/f.u. respectively). This preference stems from the large mismatch between the radii of oxygen and sulfur that result in very different ideal Zr-S and Zr-O bond lengths. A \textit{trans} order is optimal as the presence of only either Zr-S or Zr-O bonds along each pseudocubic direction leads to the smallest deviations from the optimal bond lengths. These large energy differences also indicate that the synthesis of perovskite oxysulfides with anion stoichiometries different from X\textsubscript{2}Y might be difficult. 

The relaxed structure of the mixed-anion compounds are shown in Fig. \ref{fig:unstrained} (c) and (d). While BaZrOS\textsubscript{2} still has large octahedral rotations they completely vanish in BaZrO\textsubscript{2}S and the structure develops a strong polar distortion along the Zr-S bond. This distortion is so large that it changes the local B-site coordination from octahedral to tetrahedral. We confirm the weak bond strength of the elongated Zr-S bond by COHP analysis, which shows strong bonding contributions only for the short Zr-S bond; the integrated COHP up to the Fermi level reveal that the shorter Zr-S bond is one order of magnitude stronger than the longer Zr-S bond.

\begin{figure}
	\includegraphics[width=0.9\columnwidth]{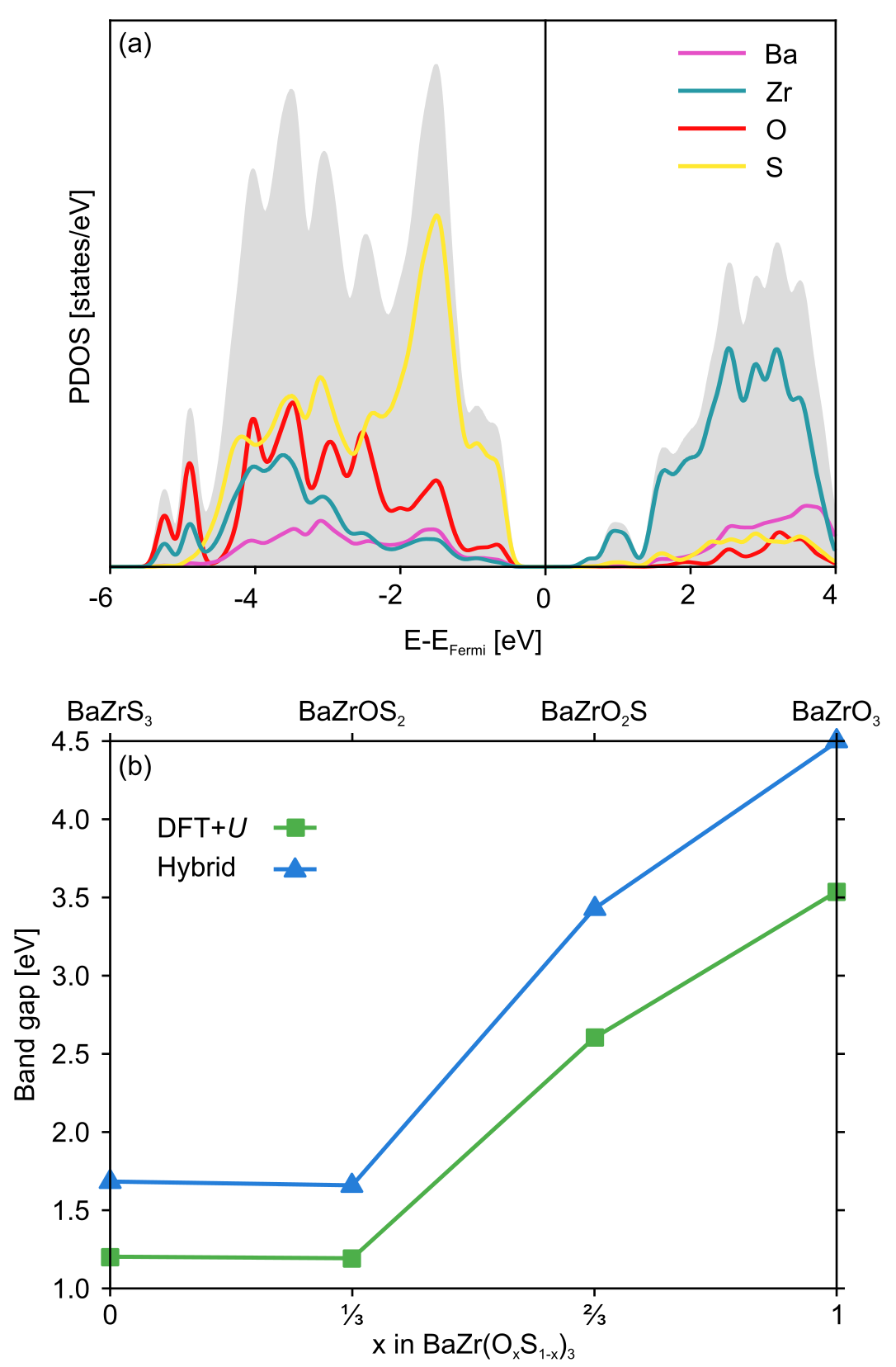}
	\caption{(a) Partial density of states of BaZrOS\textsubscript{2} at the PBE+\textit{U} level of theory and (b) PBE+\textit{U} and HSE band gaps in the BaZr(O\textsubscript{x}S\textsubscript{1-x})\textsubscript{3} series.}
	\label{fig:pdos_bandgap}
\end{figure} 

Having established the preferred anion order of the mixed-anion members, we now turn to the electronic properties of the BaZr(O\textsubscript{x}S\textsubscript{1-x})\textsubscript{3} series. The conduction-band edge is Zr 4d dominated in all compounds. The valence-band edge in the pure oxide is O 2p dominated, but due to the smaller electronegativity of sulfur it is formed by S 3p states in the oxysulfides and the sulfide. We show the projected density of states (PDOS) for BaZrOS$_2$ in Fig. \ref{fig:pdos_bandgap} (a). Fig. \ref{fig:pdos_bandgap} (b), shows the evolution of the DFT+$U$ and HSE band gaps as a function of the oxygen content. Due to the S 3p dominated valence band-edge we expect no marked change in band gap for sulfur-containing compounds. This is indeed observed for BaZrS$_3$ and BaZrOS$_2$ but for BaZrO$_2$S the band gap increases markedly. We can relate this to the emergence of the polar distortion shown in Fig. \ref{fig:ferro} (d), which was shown to drastically increase the band gap for mixed-anion compounds such as oxynitrides \cite{vonruti2018epitaxial}. The same behavior was also observed experimentally, but not further investigated \cite{perera2016chalcogenide}. While our HSE band gaps agree well with experiment for BaZrS\textsubscript{3} and BaZrOS\textsubscript{2}, the value predicted for BaZrO\textsubscript{2}S (3.4 eV) is much larger  than the experimental value of 2.5 eV \cite{perera2016chalcogenide}. This mismatch could stem from a small anion disorder in experiment that would reduce the polar distortion and hence the band gap.

\begin{figure}
	\includegraphics[width=0.9\columnwidth]{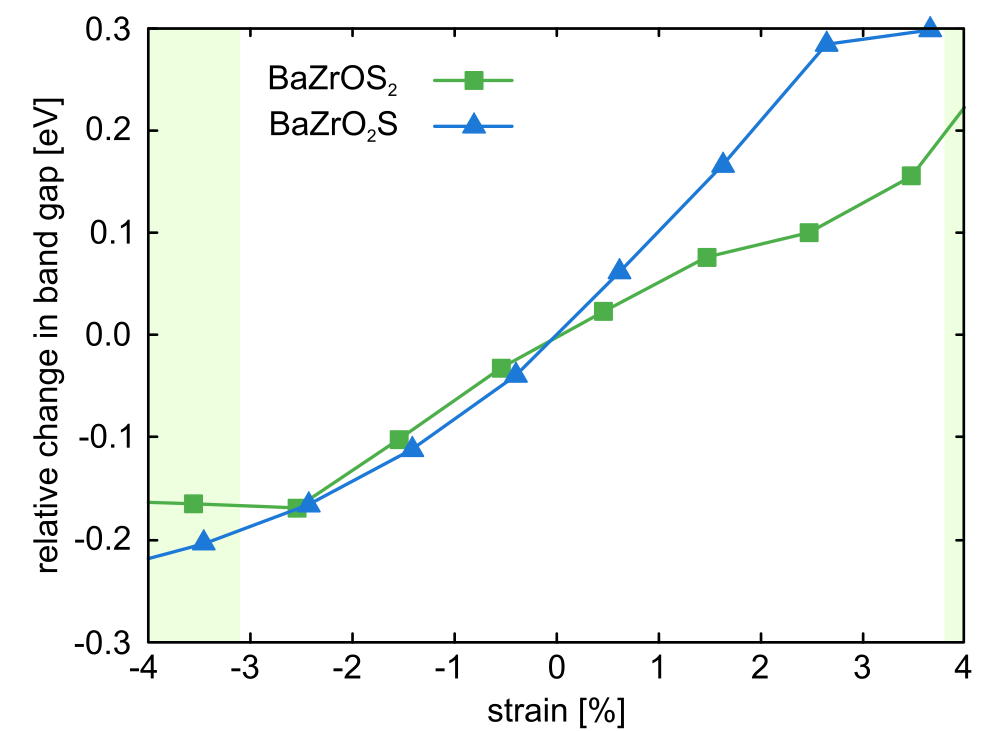}
	\caption{Relative change in band gap for BaZrOS\textsubscript{2} and BaZrO\textsubscript{2}S. The green coloured background indicates the appearance of polar distortions for BaZrOS\textsubscript{2}. BaZrO\textsubscript{2}S shows polar distortions at all investigated strains.}
	\label{fig:bandgapstrain}
\end{figure}

The HSE band gap of BaZrO\textsubscript{2}S is around 3.5 eV and would need to be reduced to improve the photocatalytic activity, whereas BaZrOS\textsubscript{2} has a HSE band gap of 1.66 eV that is too small to promote photocatalytic water splitting. We therefore apply epitaxial strain to both materials with the goal of changing the amplitude of the polar distortion and tune their band gaps to values suitable for photocatalytic water splitting. We consider only orientations of the 1D \textit{trans} chains along the film normal as preliminary test for other orientations (not shown) were energetically significantly less favorable due to the very different length of Zr-O and Zr-S bonds. As shown by the green background in Fig. \ref{fig:bandgapstrain}, we find that both large compressive and tensile strain induce polar distortion in BaZrOS\textsubscript{2}. For compressive strain the distortion is oriented along the out-of-plane direction (along Zr-O bonds) and for tensile strain along the in-plane direction (along Zr-S bonds), which is not relevant for applications in photocatalysis. Looking at the band gap (green curve in Fig. \ref{fig:bandgapstrain}), we observe in the intermediate strain range, where the structure is non-polar, a continuous decrease in the band gap with increasing compressive strain that stems from increased bandwidths with decreased volume. While we see an increase in band gap when the out-of-plane polar distortion appears beyond 3\% compressive strain, there is no overall increase in band gap for compressively strained BaZrO\textsubscript{2}S and the band gap is therefore still too small for photocatalytic water splitting. 
 
BaZrO\textsubscript{2}S has a polar distortion already without strain and we would expect strong changes in band gap due to its strain-dependent modulation. The blue curve in Fig. \ref{fig:bandgapstrain} however shows only small changes comparable to the volume-induced ones discussed above for BaZrOS\textsubscript{2}. We believe this to stem from the fact that a very large polar distortion as shown in Fig. \ref{fig:unstrained} (d) leads to changes in the band-edge forming orbitals as shown in Fig. \ref{fig:sketch_bandgap} (c). While strain does modulate the amplitude of the polar distortion, this has no longer an effect on the band gap. This leads to band gaps that are unfortunately still too high for applications in photocatalytic water splitting.

\subsection{Changing the conduction-band edge by B site alloying: BaHfOS\textsubscript{2} and BaZr\textsubscript{y}Ti\textsubscript{1-y}O\textsubscript{2}S}

\begin{figure}
	\includegraphics[width=0.9\columnwidth]{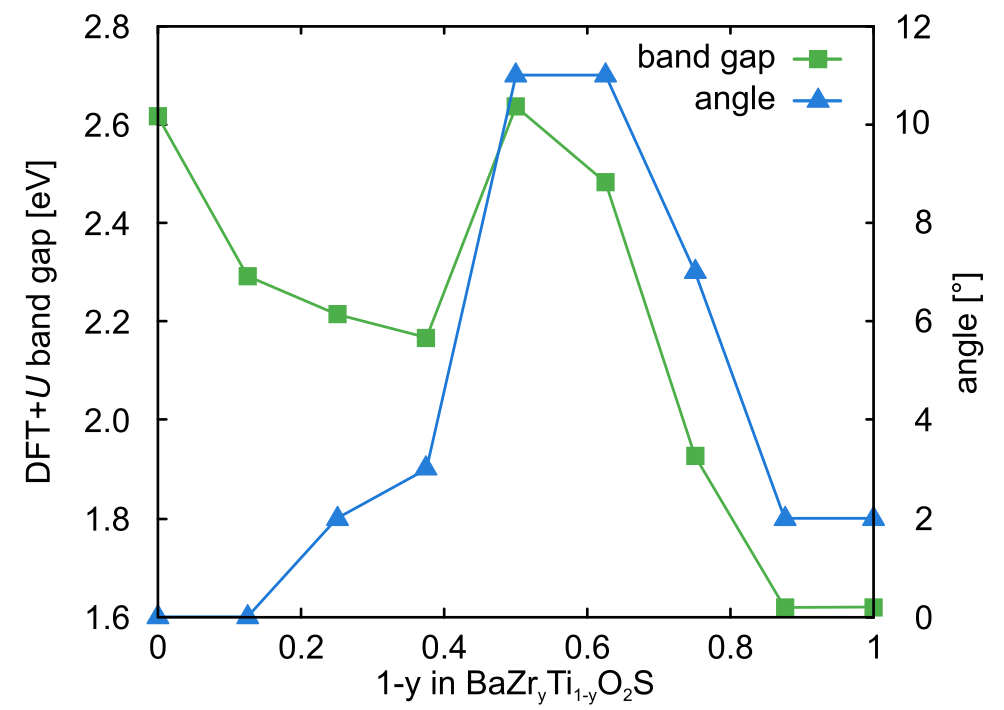}
	\caption{DFT+$U$ band gaps and tetrahedral rotation angles for the most stable BaZr\textsubscript{y}Ti\textsubscript{1-y}O\textsubscript{2}S structures.}
	\label{fig:bandgapTi}
\end{figure}

\begin{figure}
	\centering
	\includegraphics[width=0.95\columnwidth]{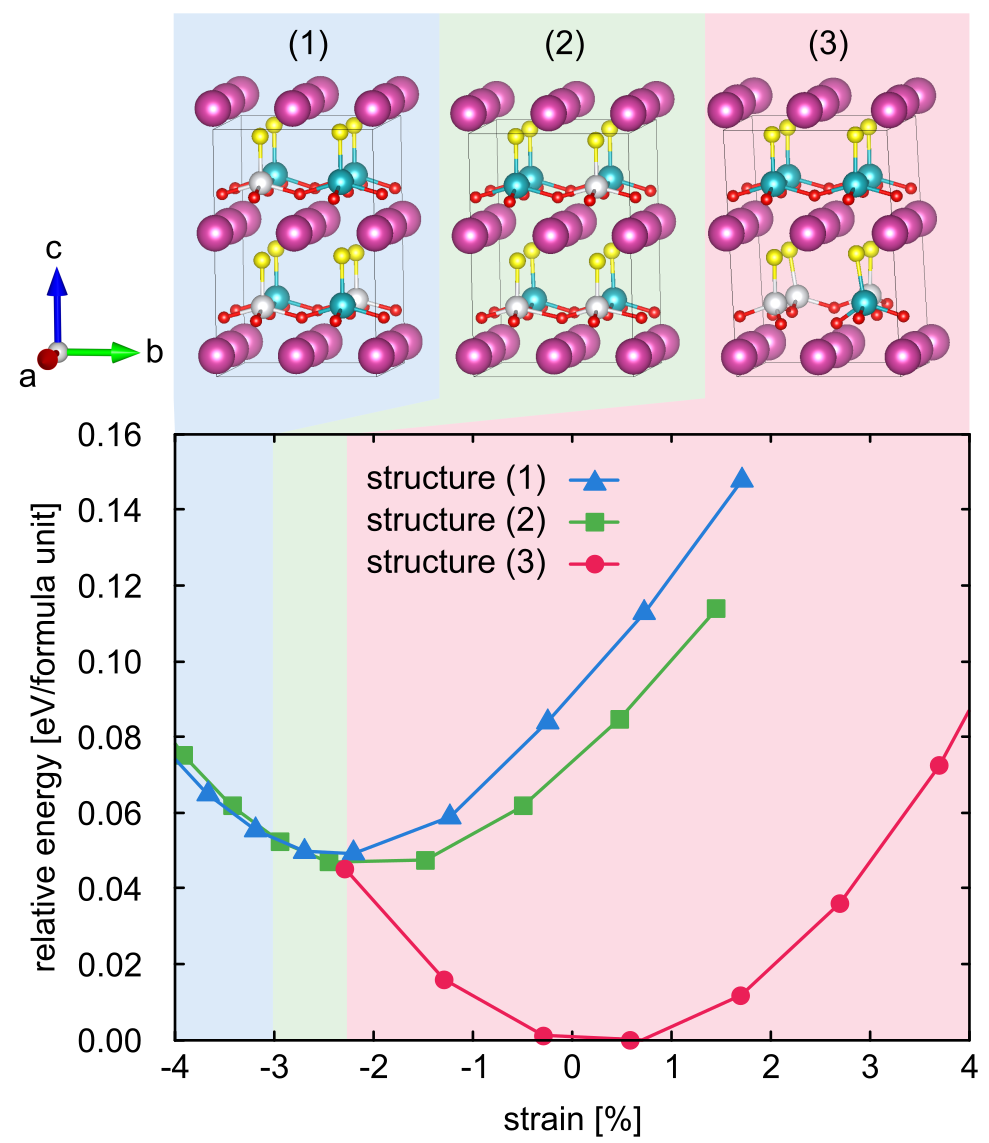}
	\caption{Relative energy of different low energy structures of BaZr\textsubscript{0.625}Ti\textsubscript{0.375}O\textsubscript{2}S: (1) no Ti clustering or polyhedra rotations, lowest energy for compressive strain larger than 3\%, (2) Ti clustering but no polyhedra rotations, lowest energy between 3 and 2.2 \% compressive strain and (3) Ti clustering and polyhedra rotations, lowest energy above 2.2\% compressive strain. Color code: Ba=purple, Zr=cyan, Ti = white, O = red, S= yellow.}
	\label{fig:Bmix}
\end{figure}

Given that band gaps for strained BaZrOS\textsubscript{2} and BaZrO\textsubscript{2}S are still too small respectively too large for photocatalytic water splitting, we change the B site to further engineer the band gap. Since BaZrOS\textsubscript{2} has a too small band gap, we substitute Zr by Hf in order to increase the band gap. As expected, BaHfOS\textsubscript{2} has a higher-lying  conduction-band edge but we do not find the emergence of polar distortions for experimentally accessible strains (not shown). For BaZrO\textsubscript{2}S on the other hand, we try to decrease the conduction-band edge and hence the band gap by substituting Zr with Ti. A full substitution of Zr by Ti leads to a large decrease in band gap ($\Delta E_{g,HSE}$ = 1.3 eV) while the structure preserves its strong polar distortion. We hence investigate partial substitution by computing for different Ti content all symmetrically distinct Zr/Ti arrangements on the 8 B sites in the 40 atom unit cell. The band gap of the most stable structure for each BaZr\textsubscript{y}Ti\textsubscript{1-y}O\textsubscript{2}S composition is shown by the green squares in Fig. \ref{fig:bandgapTi}. While the band gap generally decreases with increasing titanium content, this trend is broken for $1-y$ = 0.500 and 0.625. For these structures we find large rotations of the tetrahedra (see Fig. \ref{fig:Bmix} d) that we characterize by the deviation of the B-S bond from the $c$ axis (blue triangles in Fig. \ref{fig:bandgapTi}). This is in agreement with the fact that rotations of the coordination polyhedra, like polar distortions, lead to an increased band gap in perovskite oxides \cite{berger2011band}.
 
The observed tetrahedral rotations indicate a potential instability of the perovskite structure for $1-y>0.2$, which is supported by two additional observations: First, we observe that in the energetically most favorable structures B sites of the same element tend to cluster and second, we see a bond-length asymmetry of O-Ti-O bonds as shown for structure (2) in Fig. \ref{fig:Bmix}. We further investigate the potential to stabilise the perovskite-like BaZr\textsubscript{y}Ti\textsubscript{1-y}O\textsubscript{2}S structure by strain. This is motivated by previous findings for BaZr\textsubscript{y}Ti\textsubscript{1-y}S\textsubscript{3}, where BaTiS\textsubscript{3} was found to be unstable in the perovskite structure but the substitution of Ti by Zr and compressive biaxial strain both lead to a stabilization of the perovskite structure \cite{huster1980kristallstruktur, meng2016alloying}. As shown for BaZr\textsubscript{0.625}Ti\textsubscript{0.375}O\textsubscript{2}S in Fig. \ref{fig:Bmix}, we find indeed that compressive strain strongly destabilises structure (3), which has tetrahedral rotations and Ti-clustering. For compressive strain around 2.2\% first a structure without tetrahedral rotations becomes stable, while beyond 3\% compressive strain also the Ti-clustering vanishes. These findings are general for $1-y\le 0.375$, whereas for larger Ti content the perovskite-like structure is unstable for all investigated strains. Compressive strain also reduces the band gap and for 3\% compressive strained BaZr\textsubscript{0.675}Ti\textsubscript{0.325}O\textsubscript{2}S we predict a HSE band gap of 2.8 eV. We want to stress again that anion disorder could further reduce the band gap by a reduced polar distortion.

\subsection{Electric polarization and coercive field }

\begin{figure}
	\includegraphics[width=0.9\columnwidth]{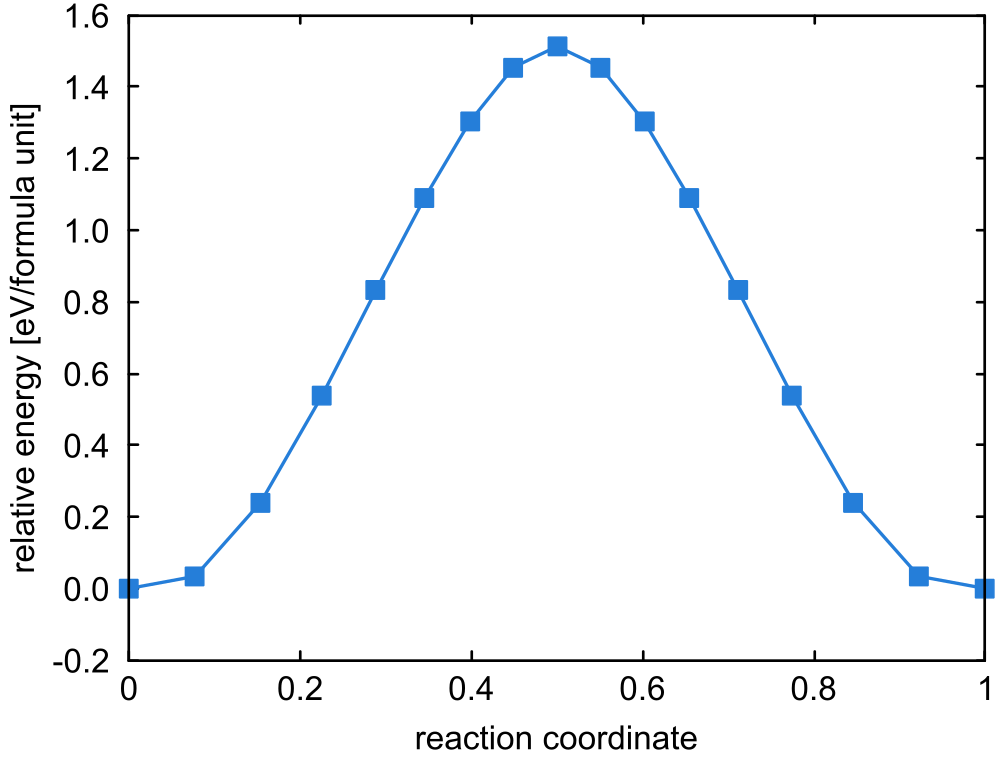}
	\caption{Energy profile for polarization switching in BaZrO\textsubscript{2}S.}
	\label{fig:switch}
\end{figure}

Finally, we determine the coercive field of BaZrO\textsubscript{2}S, which is defined as
\begin{equation}
\epsilon_c = \frac{(4/3)^{(3/2)}E}{P \cdot V},
\label{eq:polarization}
\end{equation}
where $E$ is the switching barrier per formula unit between the two oppositely polarized states, $P$ the electric polarization at the bottom of the double well potential normalized per area perpendicular to the polarization direction and $V$ the volume per formula unit \cite{beckman2009ideal}. We determine the switching barrier to be 1.5 eV/formula unit (see Fig. \ref{fig:switch}) via a nudged elastic band calculation and the polarisation to be 0.88 C/m\textsuperscript{2} via the Berry phase method, using the transition state as a non-polar reference to determine the quantum of polarization. While the polarization is slightly larger compared to traditional ferroelectrics such as PbTiO\textsubscript{3} \cite{beckman2009ideal}, the switching energy is significantly larger and results in a coercive field of 2143 MV/m, one order of magnitude larger than in PbTiO\textsubscript{3} \cite{beckman2009ideal}. Substitution of the A site with Ca and Zr as well as the B site with Ti do not result in significantly smaller coercive fields (not shown). We therefore conclude that sulfides and oxysulfides, where the polarisation occurs along S-B-S bonds, are only polar but not ferroelectric due to their prohibitively large coercive fields.

\section{Conclusion}

We investigated the AB(O\textsubscript{x}S\textsubscript{1-x})\textsubscript{3} system as polar photocatalysts for water splitting. We find that oxysulfides strongly prefer a $trans$ anion configuration due to the large size mismatch of oxygen and sulfur, which is contrary to most mixed-anion perovskites that favor a $cis$ configuration. The $trans$ anion configuration in ABO\textsubscript{2}S oxysulfide thin films is directly related to the appearance of a strong polar distortion along the technologically relevant out-of-plane direction. Due to the high switching barrier, these materials are however only polar and not ferroelectric.

Increasing the Ti content in the BaZr\textsubscript{y}Ti\textsubscript{1-y}O\textsubscript{2}S series lowers the band gap but also leads to structural instabilities that can be stabilized by compressive strain up to $1-y$ = 0.375. We find 3\% compressively strained BaZr\textsubscript{0.625}Ti\textsubscript{0.375}O\textsubscript{2}S to have the most promising 2.8 eV HSE band gap for photocatalytic water splitting. Comparison with experiment for the BaZr(O\textsubscript{x}S\textsubscript{1-x})\textsubscript{3} series suggests that band gaps could be smaller than our HSE estimates due anion disorder. Mixed cation oxysulfides are therefore promising candidates for highly efficient water-splitting photocatalysts that combine strong polarity with small band gaps. 
 
\section{Acknowledgements}

This research was funded by the SNF Professorship Grant PP00P2\_157615. Calculations were performed on UBELIX (http://www.id.unibe.ch/hpc), the HPC cluster at the University of Bern as well as the Swiss National Supercomputing Centre (CSCS) under project ID 766 and SuperMUC at GCS@LRZ, Germany, for which we acknowledge PRACE for awarding us access.

\bibliography{library}

\begin{thebibliography}{49}%
\makeatletter
\providecommand \@ifxundefined [1]{%
 \@ifx{#1\undefined}
}%
\providecommand \@ifnum [1]{%
 \ifnum #1\expandafter \@firstoftwo
 \else \expandafter \@secondoftwo
 \fi
}%
\providecommand \@ifx [1]{%
 \ifx #1\expandafter \@firstoftwo
 \else \expandafter \@secondoftwo
 \fi
}%
\providecommand \natexlab [1]{#1}%
\providecommand \enquote  [1]{``#1''}%
\providecommand \bibnamefont  [1]{#1}%
\providecommand \bibfnamefont [1]{#1}%
\providecommand \citenamefont [1]{#1}%
\providecommand \href@noop [0]{\@secondoftwo}%
\providecommand \href [0]{\begingroup \@sanitize@url \@href}%
\providecommand \@href[1]{\@@startlink{#1}\@@href}%
\providecommand \@@href[1]{\endgroup#1\@@endlink}%
\providecommand \@sanitize@url [0]{\catcode `\\12\catcode `\$12\catcode
  `\&12\catcode `\#12\catcode `\^12\catcode `\_12\catcode `\%12\relax}%
\providecommand \@@startlink[1]{}%
\providecommand \@@endlink[0]{}%
\providecommand \url  [0]{\begingroup\@sanitize@url \@url }%
\providecommand \@url [1]{\endgroup\@href {#1}{\urlprefix }}%
\providecommand \urlprefix  [0]{URL }%
\providecommand \Eprint [0]{\href }%
\providecommand \doibase [0]{http://dx.doi.org/}%
\providecommand \selectlanguage [0]{\@gobble}%
\providecommand \bibinfo  [0]{\@secondoftwo}%
\providecommand \bibfield  [0]{\@secondoftwo}%
\providecommand \translation [1]{[#1]}%
\providecommand \BibitemOpen [0]{}%
\providecommand \bibitemStop [0]{}%
\providecommand \bibitemNoStop [0]{.\EOS\space}%
\providecommand \EOS [0]{\spacefactor3000\relax}%
\providecommand \BibitemShut  [1]{\csname bibitem#1\endcsname}%
\let\auto@bib@innerbib\@empty
\bibitem [{\citenamefont {Giocondi}\ and\ \citenamefont
  {Rohrer}(2001)}]{giocondi2001spatial}%
  \BibitemOpen
  \bibfield  {author} {\bibinfo {author} {\bibfnamefont {J.~L.}\ \bibnamefont
  {Giocondi}}\ and\ \bibinfo {author} {\bibfnamefont {G.~S.}\ \bibnamefont
  {Rohrer}},\ }\href@noop {} {\bibfield  {journal} {\bibinfo  {journal} {The
  Journal of Physical Chemistry B}\ }\textbf {\bibinfo {volume} {105}},\
  \bibinfo {pages} {8275} (\bibinfo {year} {2001})}\BibitemShut {NoStop}%
\bibitem [{\citenamefont {Li}\ \emph {et~al.}(2014)\citenamefont {Li},
  \citenamefont {Salvador},\ and\ \citenamefont
  {Rohrer}}]{li2014photocatalysts}%
  \BibitemOpen
  \bibfield  {author} {\bibinfo {author} {\bibfnamefont {L.}~\bibnamefont
  {Li}}, \bibinfo {author} {\bibfnamefont {P.~A.}\ \bibnamefont {Salvador}}, \
  and\ \bibinfo {author} {\bibfnamefont {G.~S.}\ \bibnamefont {Rohrer}},\
  }\href@noop {} {\bibfield  {journal} {\bibinfo  {journal} {Nanoscale}\
  }\textbf {\bibinfo {volume} {6}},\ \bibinfo {pages} {24} (\bibinfo {year}
  {2014})}\BibitemShut {NoStop}%
\bibitem [{\citenamefont {Matthias}(1949)}]{matthias1949new}%
  \BibitemOpen
  \bibfield  {author} {\bibinfo {author} {\bibfnamefont {B.}~\bibnamefont
  {Matthias}},\ }\href@noop {} {\bibfield  {journal} {\bibinfo  {journal}
  {Physical Review}\ }\textbf {\bibinfo {volume} {75}},\ \bibinfo {pages}
  {1771} (\bibinfo {year} {1949})}\BibitemShut {NoStop}%
\bibitem [{\citenamefont {Hill}(2000)}]{hill2000there}%
  \BibitemOpen
  \bibfield  {author} {\bibinfo {author} {\bibfnamefont {N.~A.}\ \bibnamefont
  {Hill}},\ }\href@noop {} {\bibfield  {journal} {\bibinfo  {journal} {The
  Journal of Physical Chemistry B}\ }\textbf {\bibinfo {volume} {104}},\
  \bibinfo {pages} {6694} (\bibinfo {year} {2000})}\BibitemShut {NoStop}%
\bibitem [{\citenamefont {Bolton}\ \emph {et~al.}(1985)\citenamefont {Bolton},
  \citenamefont {Strickler},\ and\ \citenamefont
  {Connolly}}]{bolton1985limiting}%
  \BibitemOpen
  \bibfield  {author} {\bibinfo {author} {\bibfnamefont {J.~R.}\ \bibnamefont
  {Bolton}}, \bibinfo {author} {\bibfnamefont {S.~J.}\ \bibnamefont
  {Strickler}}, \ and\ \bibinfo {author} {\bibfnamefont {J.~S.}\ \bibnamefont
  {Connolly}},\ }\href@noop {} {\bibfield  {journal} {\bibinfo  {journal}
  {Nature}\ }\textbf {\bibinfo {volume} {316}},\ \bibinfo {pages} {495}
  (\bibinfo {year} {1985})}\BibitemShut {NoStop}%
\bibitem [{\citenamefont {Kasahara}\ \emph {et~al.}(2002)\citenamefont
  {Kasahara}, \citenamefont {Nukumizu}, \citenamefont {Hitoki}, \citenamefont
  {Takata}, \citenamefont {Kondo}, \citenamefont {Hara}, \citenamefont
  {Kobayashi},\ and\ \citenamefont {Domen}}]{kasahara2002photoreactions}%
  \BibitemOpen
  \bibfield  {author} {\bibinfo {author} {\bibfnamefont {A.}~\bibnamefont
  {Kasahara}}, \bibinfo {author} {\bibfnamefont {K.}~\bibnamefont {Nukumizu}},
  \bibinfo {author} {\bibfnamefont {G.}~\bibnamefont {Hitoki}}, \bibinfo
  {author} {\bibfnamefont {T.}~\bibnamefont {Takata}}, \bibinfo {author}
  {\bibfnamefont {J.~N.}\ \bibnamefont {Kondo}}, \bibinfo {author}
  {\bibfnamefont {M.}~\bibnamefont {Hara}}, \bibinfo {author} {\bibfnamefont
  {H.}~\bibnamefont {Kobayashi}}, \ and\ \bibinfo {author} {\bibfnamefont
  {K.}~\bibnamefont {Domen}},\ }\href@noop {} {\bibfield  {journal} {\bibinfo
  {journal} {The Journal of Physical Chemistry A}\ }\textbf {\bibinfo {volume}
  {106}},\ \bibinfo {pages} {6750} (\bibinfo {year} {2002})}\BibitemShut
  {NoStop}%
\bibitem [{\citenamefont {Oka}\ \emph {et~al.}(2014)\citenamefont {Oka},
  \citenamefont {Hirose}, \citenamefont {Kamisaka}, \citenamefont {Fukumura},
  \citenamefont {Sasa}, \citenamefont {Ishii}, \citenamefont {Matsuzaki},
  \citenamefont {Sato}, \citenamefont {Ikuhara},\ and\ \citenamefont
  {Hasegawa}}]{oka2014possible}%
  \BibitemOpen
  \bibfield  {author} {\bibinfo {author} {\bibfnamefont {D.}~\bibnamefont
  {Oka}}, \bibinfo {author} {\bibfnamefont {Y.}~\bibnamefont {Hirose}},
  \bibinfo {author} {\bibfnamefont {H.}~\bibnamefont {Kamisaka}}, \bibinfo
  {author} {\bibfnamefont {T.}~\bibnamefont {Fukumura}}, \bibinfo {author}
  {\bibfnamefont {K.}~\bibnamefont {Sasa}}, \bibinfo {author} {\bibfnamefont
  {S.}~\bibnamefont {Ishii}}, \bibinfo {author} {\bibfnamefont
  {H.}~\bibnamefont {Matsuzaki}}, \bibinfo {author} {\bibfnamefont
  {Y.}~\bibnamefont {Sato}}, \bibinfo {author} {\bibfnamefont {Y.}~\bibnamefont
  {Ikuhara}}, \ and\ \bibinfo {author} {\bibfnamefont {T.}~\bibnamefont
  {Hasegawa}},\ }\href@noop {} {\bibfield  {journal} {\bibinfo  {journal}
  {Scientific Reports}\ }\textbf {\bibinfo {volume} {4}},\ \bibinfo {pages}
  {4987} (\bibinfo {year} {2014})}\BibitemShut {NoStop}%
\bibitem [{\citenamefont {Vonr{\"u}ti}\ and\ \citenamefont
  {Aschauer}(2018{\natexlab{a}})}]{vonruti2018anion}%
  \BibitemOpen
  \bibfield  {author} {\bibinfo {author} {\bibfnamefont {N.}~\bibnamefont
  {Vonr{\"u}ti}}\ and\ \bibinfo {author} {\bibfnamefont {U.}~\bibnamefont
  {Aschauer}},\ }\href@noop {} {\bibfield  {journal} {\bibinfo  {journal}
  {Physical Review Letters}\ }\textbf {\bibinfo {volume} {120}},\ \bibinfo
  {pages} {046001} (\bibinfo {year} {2018}{\natexlab{a}})}\BibitemShut
  {NoStop}%
\bibitem [{\citenamefont {Bersuker}(2013)}]{bersuker2013pseudo}%
  \BibitemOpen
  \bibfield  {author} {\bibinfo {author} {\bibfnamefont {I.~B.}\ \bibnamefont
  {Bersuker}},\ }\href@noop {} {\bibfield  {journal} {\bibinfo  {journal}
  {Chemical Reviews}\ }\textbf {\bibinfo {volume} {113}},\ \bibinfo {pages}
  {1351} (\bibinfo {year} {2013})}\BibitemShut {NoStop}%
\bibitem [{\citenamefont {Rondinelli}\ \emph {et~al.}(2009)\citenamefont
  {Rondinelli}, \citenamefont {Eidelson},\ and\ \citenamefont
  {Spaldin}}]{rondinelli2009non}%
  \BibitemOpen
  \bibfield  {author} {\bibinfo {author} {\bibfnamefont {J.~M.}\ \bibnamefont
  {Rondinelli}}, \bibinfo {author} {\bibfnamefont {A.~S.}\ \bibnamefont
  {Eidelson}}, \ and\ \bibinfo {author} {\bibfnamefont {N.~A.}\ \bibnamefont
  {Spaldin}},\ }\href@noop {} {\bibfield  {journal} {\bibinfo  {journal}
  {Physical Review B}\ }\textbf {\bibinfo {volume} {79}},\ \bibinfo {pages}
  {205119} (\bibinfo {year} {2009})}\BibitemShut {NoStop}%
\bibitem [{\citenamefont {Wheeler}\ \emph {et~al.}(1986)\citenamefont
  {Wheeler}, \citenamefont {Whangbo}, \citenamefont {Hughbanks}, \citenamefont
  {Hoffmann}, \citenamefont {Burdett},\ and\ \citenamefont
  {Albright}}]{wheeler1986symmetric}%
  \BibitemOpen
  \bibfield  {author} {\bibinfo {author} {\bibfnamefont {R.~A.}\ \bibnamefont
  {Wheeler}}, \bibinfo {author} {\bibfnamefont {M.~H.}\ \bibnamefont
  {Whangbo}}, \bibinfo {author} {\bibfnamefont {T.}~\bibnamefont {Hughbanks}},
  \bibinfo {author} {\bibfnamefont {R.}~\bibnamefont {Hoffmann}}, \bibinfo
  {author} {\bibfnamefont {J.~K.}\ \bibnamefont {Burdett}}, \ and\ \bibinfo
  {author} {\bibfnamefont {T.~A.}\ \bibnamefont {Albright}},\ }\href@noop {}
  {\bibfield  {journal} {\bibinfo  {journal} {Journal of the American Chemical
  Society}\ }\textbf {\bibinfo {volume} {108}},\ \bibinfo {pages} {2222}
  (\bibinfo {year} {1986})}\BibitemShut {NoStop}%
\bibitem [{\citenamefont {Vonr{\"u}ti}\ and\ \citenamefont
  {Aschauer}(2018{\natexlab{b}})}]{vonruti2018epitaxial}%
  \BibitemOpen
  \bibfield  {author} {\bibinfo {author} {\bibfnamefont {N.}~\bibnamefont
  {Vonr{\"u}ti}}\ and\ \bibinfo {author} {\bibfnamefont {U.}~\bibnamefont
  {Aschauer}},\ }\href@noop {} {\bibfield  {journal} {\bibinfo  {journal}
  {Physical Review Materials}\ }\textbf {\bibinfo {volume} {2}},\ \bibinfo
  {pages} {105401} (\bibinfo {year} {2018}{\natexlab{b}})}\BibitemShut
  {NoStop}%
\bibitem [{\citenamefont {Berger}\ \emph {et~al.}(2011)\citenamefont {Berger},
  \citenamefont {Fennie},\ and\ \citenamefont {Neaton}}]{berger2011band}%
  \BibitemOpen
  \bibfield  {author} {\bibinfo {author} {\bibfnamefont {R.~F.}\ \bibnamefont
  {Berger}}, \bibinfo {author} {\bibfnamefont {C.~J.}\ \bibnamefont {Fennie}},
  \ and\ \bibinfo {author} {\bibfnamefont {J.~B.}\ \bibnamefont {Neaton}},\
  }\href@noop {} {\bibfield  {journal} {\bibinfo  {journal} {Physical Review
  Letters}\ }\textbf {\bibinfo {volume} {107}},\ \bibinfo {pages} {146804}
  (\bibinfo {year} {2011})}\BibitemShut {NoStop}%
\bibitem [{\citenamefont {Sun}\ \emph {et~al.}(2015)\citenamefont {Sun},
  \citenamefont {Agiorgousis}, \citenamefont {Zhang},\ and\ \citenamefont
  {Zhang}}]{sun2015chalcogenide}%
  \BibitemOpen
  \bibfield  {author} {\bibinfo {author} {\bibfnamefont {Y.-Y.}\ \bibnamefont
  {Sun}}, \bibinfo {author} {\bibfnamefont {M.~L.}\ \bibnamefont
  {Agiorgousis}}, \bibinfo {author} {\bibfnamefont {P.}~\bibnamefont {Zhang}},
  \ and\ \bibinfo {author} {\bibfnamefont {S.}~\bibnamefont {Zhang}},\
  }\href@noop {} {\bibfield  {journal} {\bibinfo  {journal} {Nano Letters}\
  }\textbf {\bibinfo {volume} {15}},\ \bibinfo {pages} {581} (\bibinfo {year}
  {2015})}\BibitemShut {NoStop}%
\bibitem [{\citenamefont {Ju}\ \emph {et~al.}(2017)\citenamefont {Ju},
  \citenamefont {Dai}, \citenamefont {Ma},\ and\ \citenamefont
  {Zeng}}]{ju2017perovskite}%
  \BibitemOpen
  \bibfield  {author} {\bibinfo {author} {\bibfnamefont {M.-G.}\ \bibnamefont
  {Ju}}, \bibinfo {author} {\bibfnamefont {J.}~\bibnamefont {Dai}}, \bibinfo
  {author} {\bibfnamefont {L.}~\bibnamefont {Ma}}, \ and\ \bibinfo {author}
  {\bibfnamefont {X.~C.}\ \bibnamefont {Zeng}},\ }\href@noop {} {\bibfield
  {journal} {\bibinfo  {journal} {Advanced Energy Materials}\ }\textbf
  {\bibinfo {volume} {7}},\ \bibinfo {pages} {1700216} (\bibinfo {year}
  {2017})}\BibitemShut {NoStop}%
\bibitem [{\citenamefont {Sun}\ \emph {et~al.}(2018)\citenamefont {Sun},
  \citenamefont {Hisatomi}, \citenamefont {Wang}, \citenamefont {Chen},
  \citenamefont {Ma}, \citenamefont {Liu}, \citenamefont {Nandy}, \citenamefont
  {Minegishi}, \citenamefont {Katayama},\ and\ \citenamefont
  {Domen}}]{sun2018efficient}%
  \BibitemOpen
  \bibfield  {author} {\bibinfo {author} {\bibfnamefont {S.}~\bibnamefont
  {Sun}}, \bibinfo {author} {\bibfnamefont {T.}~\bibnamefont {Hisatomi}},
  \bibinfo {author} {\bibfnamefont {Q.}~\bibnamefont {Wang}}, \bibinfo {author}
  {\bibfnamefont {S.}~\bibnamefont {Chen}}, \bibinfo {author} {\bibfnamefont
  {G.}~\bibnamefont {Ma}}, \bibinfo {author} {\bibfnamefont {J.}~\bibnamefont
  {Liu}}, \bibinfo {author} {\bibfnamefont {S.}~\bibnamefont {Nandy}}, \bibinfo
  {author} {\bibfnamefont {T.}~\bibnamefont {Minegishi}}, \bibinfo {author}
  {\bibfnamefont {M.}~\bibnamefont {Katayama}}, \ and\ \bibinfo {author}
  {\bibfnamefont {K.}~\bibnamefont {Domen}},\ }\href@noop {} {\bibfield
  {journal} {\bibinfo  {journal} {ACS Catalysis}\ }\textbf {\bibinfo {volume}
  {8}},\ \bibinfo {pages} {1690} (\bibinfo {year} {2018})}\BibitemShut
  {NoStop}%
\bibitem [{\citenamefont {Ma}\ \emph {et~al.}(2016)\citenamefont {Ma},
  \citenamefont {Chen}, \citenamefont {Kuang}, \citenamefont {Akiyama},
  \citenamefont {Hisatomi}, \citenamefont {Nakabayashi}, \citenamefont
  {Shibata}, \citenamefont {Katayama}, \citenamefont {Minegishi},\ and\
  \citenamefont {Domen}}]{ma2016visible}%
  \BibitemOpen
  \bibfield  {author} {\bibinfo {author} {\bibfnamefont {G.}~\bibnamefont
  {Ma}}, \bibinfo {author} {\bibfnamefont {S.}~\bibnamefont {Chen}}, \bibinfo
  {author} {\bibfnamefont {Y.}~\bibnamefont {Kuang}}, \bibinfo {author}
  {\bibfnamefont {S.}~\bibnamefont {Akiyama}}, \bibinfo {author} {\bibfnamefont
  {T.}~\bibnamefont {Hisatomi}}, \bibinfo {author} {\bibfnamefont
  {M.}~\bibnamefont {Nakabayashi}}, \bibinfo {author} {\bibfnamefont
  {N.}~\bibnamefont {Shibata}}, \bibinfo {author} {\bibfnamefont
  {M.}~\bibnamefont {Katayama}}, \bibinfo {author} {\bibfnamefont
  {T.}~\bibnamefont {Minegishi}}, \ and\ \bibinfo {author} {\bibfnamefont
  {K.}~\bibnamefont {Domen}},\ }\href@noop {} {\bibfield  {journal} {\bibinfo
  {journal} {The Journal of Physical Chemistry Letters}\ }\textbf {\bibinfo
  {volume} {7}},\ \bibinfo {pages} {3892} (\bibinfo {year} {2016})}\BibitemShut
  {NoStop}%
\bibitem [{\citenamefont {Shannon}(1976)}]{shannon1976revised}%
  \BibitemOpen
  \bibfield  {author} {\bibinfo {author} {\bibfnamefont {R.~D.}\ \bibnamefont
  {Shannon}},\ }\href@noop {} {\bibfield  {journal} {\bibinfo  {journal} {Acta
  Crystallographica Section A: Crystal Physics, Diffraction, Theoretical and
  General Crystallography}\ }\textbf {\bibinfo {volume} {32}},\ \bibinfo
  {pages} {751} (\bibinfo {year} {1976})}\BibitemShut {NoStop}%
\bibitem [{\citenamefont {Lelieveld}\ and\ \citenamefont
  {Ijdo}(1980)}]{lelieveld1980sulphides}%
  \BibitemOpen
  \bibfield  {author} {\bibinfo {author} {\bibfnamefont {R.}~\bibnamefont
  {Lelieveld}}\ and\ \bibinfo {author} {\bibfnamefont {D.}~\bibnamefont
  {Ijdo}},\ }\href@noop {} {\bibfield  {journal} {\bibinfo  {journal} {Acta
  Crystallographica Section B}\ }\textbf {\bibinfo {volume} {36}},\ \bibinfo
  {pages} {2223} (\bibinfo {year} {1980})}\BibitemShut {NoStop}%
\bibitem [{\citenamefont {Kobayashi}\ \emph {et~al.}(2018)\citenamefont
  {Kobayashi}, \citenamefont {Tsujimoto},\ and\ \citenamefont
  {Kageyama}}]{kobayashi2018property}%
  \BibitemOpen
  \bibfield  {author} {\bibinfo {author} {\bibfnamefont {Y.}~\bibnamefont
  {Kobayashi}}, \bibinfo {author} {\bibfnamefont {Y.}~\bibnamefont
  {Tsujimoto}}, \ and\ \bibinfo {author} {\bibfnamefont {H.}~\bibnamefont
  {Kageyama}},\ }\href@noop {} {\bibfield  {journal} {\bibinfo  {journal}
  {Annual Review of Materials Research}\ }\textbf {\bibinfo {volume} {48}},\
  \bibinfo {pages} {303} (\bibinfo {year} {2018})}\BibitemShut {NoStop}%
\bibitem [{\citenamefont {Perera}\ \emph {et~al.}(2016)\citenamefont {Perera},
  \citenamefont {Hui}, \citenamefont {Zhao}, \citenamefont {Xue}, \citenamefont
  {Sun}, \citenamefont {Deng}, \citenamefont {Gross}, \citenamefont
  {Milleville}, \citenamefont {Xu}, \citenamefont {Watson} \emph
  {et~al.}}]{perera2016chalcogenide}%
  \BibitemOpen
  \bibfield  {author} {\bibinfo {author} {\bibfnamefont {S.}~\bibnamefont
  {Perera}}, \bibinfo {author} {\bibfnamefont {H.}~\bibnamefont {Hui}},
  \bibinfo {author} {\bibfnamefont {C.}~\bibnamefont {Zhao}}, \bibinfo {author}
  {\bibfnamefont {H.}~\bibnamefont {Xue}}, \bibinfo {author} {\bibfnamefont
  {F.}~\bibnamefont {Sun}}, \bibinfo {author} {\bibfnamefont {C.}~\bibnamefont
  {Deng}}, \bibinfo {author} {\bibfnamefont {N.}~\bibnamefont {Gross}},
  \bibinfo {author} {\bibfnamefont {C.}~\bibnamefont {Milleville}}, \bibinfo
  {author} {\bibfnamefont {X.}~\bibnamefont {Xu}}, \bibinfo {author}
  {\bibfnamefont {D.~F.}\ \bibnamefont {Watson}},  \emph {et~al.},\ }\href@noop
  {} {\bibfield  {journal} {\bibinfo  {journal} {Nano Energy}\ }\textbf
  {\bibinfo {volume} {22}},\ \bibinfo {pages} {129} (\bibinfo {year}
  {2016})}\BibitemShut {NoStop}%
\bibitem [{\citenamefont {Niu}\ \emph {et~al.}(2017)\citenamefont {Niu},
  \citenamefont {Huyan}, \citenamefont {Liu}, \citenamefont {Yeung},
  \citenamefont {Ye}, \citenamefont {Blankemeier}, \citenamefont {Orvis},
  \citenamefont {Sarkar}, \citenamefont {Singh}, \citenamefont {Kapadia} \emph
  {et~al.}}]{niu2017bandgap}%
  \BibitemOpen
  \bibfield  {author} {\bibinfo {author} {\bibfnamefont {S.}~\bibnamefont
  {Niu}}, \bibinfo {author} {\bibfnamefont {H.}~\bibnamefont {Huyan}}, \bibinfo
  {author} {\bibfnamefont {Y.}~\bibnamefont {Liu}}, \bibinfo {author}
  {\bibfnamefont {M.}~\bibnamefont {Yeung}}, \bibinfo {author} {\bibfnamefont
  {K.}~\bibnamefont {Ye}}, \bibinfo {author} {\bibfnamefont {L.}~\bibnamefont
  {Blankemeier}}, \bibinfo {author} {\bibfnamefont {T.}~\bibnamefont {Orvis}},
  \bibinfo {author} {\bibfnamefont {D.}~\bibnamefont {Sarkar}}, \bibinfo
  {author} {\bibfnamefont {D.~J.}\ \bibnamefont {Singh}}, \bibinfo {author}
  {\bibfnamefont {R.}~\bibnamefont {Kapadia}},  \emph {et~al.},\ }\href@noop {}
  {\bibfield  {journal} {\bibinfo  {journal} {Advanced Materials}\ }\textbf
  {\bibinfo {volume} {29}},\ \bibinfo {pages} {1604733} (\bibinfo {year}
  {2017})}\BibitemShut {NoStop}%
\bibitem [{\citenamefont {Bennett}\ \emph {et~al.}(2009)\citenamefont
  {Bennett}, \citenamefont {Grinberg},\ and\ \citenamefont
  {Rappe}}]{bennett2009effect}%
  \BibitemOpen
  \bibfield  {author} {\bibinfo {author} {\bibfnamefont {J.~W.}\ \bibnamefont
  {Bennett}}, \bibinfo {author} {\bibfnamefont {I.}~\bibnamefont {Grinberg}}, \
  and\ \bibinfo {author} {\bibfnamefont {A.~M.}\ \bibnamefont {Rappe}},\
  }\href@noop {} {\bibfield  {journal} {\bibinfo  {journal} {Physical Review
  B}\ }\textbf {\bibinfo {volume} {79}},\ \bibinfo {pages} {235115} (\bibinfo
  {year} {2009})}\BibitemShut {NoStop}%
\bibitem [{\citenamefont {Meng}\ \emph {et~al.}(2016)\citenamefont {Meng},
  \citenamefont {Saparov}, \citenamefont {Hong}, \citenamefont {Wang},
  \citenamefont {Mitzi},\ and\ \citenamefont {Yan}}]{meng2016alloying}%
  \BibitemOpen
  \bibfield  {author} {\bibinfo {author} {\bibfnamefont {W.}~\bibnamefont
  {Meng}}, \bibinfo {author} {\bibfnamefont {B.}~\bibnamefont {Saparov}},
  \bibinfo {author} {\bibfnamefont {F.}~\bibnamefont {Hong}}, \bibinfo {author}
  {\bibfnamefont {J.}~\bibnamefont {Wang}}, \bibinfo {author} {\bibfnamefont
  {D.~B.}\ \bibnamefont {Mitzi}}, \ and\ \bibinfo {author} {\bibfnamefont
  {Y.}~\bibnamefont {Yan}},\ }\href@noop {} {\bibfield  {journal} {\bibinfo
  {journal} {Chemistry of Materials}\ }\textbf {\bibinfo {volume} {28}},\
  \bibinfo {pages} {821} (\bibinfo {year} {2016})}\BibitemShut {NoStop}%
\bibitem [{\citenamefont {Brehm}\ \emph {et~al.}(2014)\citenamefont {Brehm},
  \citenamefont {Bennett}, \citenamefont {Schoenberg}, \citenamefont
  {Grinberg},\ and\ \citenamefont {Rappe}}]{brehm2014structural}%
  \BibitemOpen
  \bibfield  {author} {\bibinfo {author} {\bibfnamefont {J.~A.}\ \bibnamefont
  {Brehm}}, \bibinfo {author} {\bibfnamefont {J.~W.}\ \bibnamefont {Bennett}},
  \bibinfo {author} {\bibfnamefont {M.~R.}\ \bibnamefont {Schoenberg}},
  \bibinfo {author} {\bibfnamefont {I.}~\bibnamefont {Grinberg}}, \ and\
  \bibinfo {author} {\bibfnamefont {A.~M.}\ \bibnamefont {Rappe}},\ }\href@noop
  {} {\bibfield  {journal} {\bibinfo  {journal} {The Journal of Chemical
  Physics}\ }\textbf {\bibinfo {volume} {140}},\ \bibinfo {pages} {224703}
  (\bibinfo {year} {2014})}\BibitemShut {NoStop}%
\bibitem [{\citenamefont {Giannozzi}\ \emph {et~al.}(2009)\citenamefont
  {Giannozzi}, \citenamefont {Baroni}, \citenamefont {Bonini}, \citenamefont
  {Calandra}, \citenamefont {Car}, \citenamefont {Cavazzoni}, \citenamefont
  {Ceresoli}, \citenamefont {Chiarotti}, \citenamefont {Cococcioni},
  \citenamefont {Dabo} \emph {et~al.}}]{giannozzi2009quantum}%
  \BibitemOpen
  \bibfield  {author} {\bibinfo {author} {\bibfnamefont {P.}~\bibnamefont
  {Giannozzi}}, \bibinfo {author} {\bibfnamefont {S.}~\bibnamefont {Baroni}},
  \bibinfo {author} {\bibfnamefont {N.}~\bibnamefont {Bonini}}, \bibinfo
  {author} {\bibfnamefont {M.}~\bibnamefont {Calandra}}, \bibinfo {author}
  {\bibfnamefont {R.}~\bibnamefont {Car}}, \bibinfo {author} {\bibfnamefont
  {C.}~\bibnamefont {Cavazzoni}}, \bibinfo {author} {\bibfnamefont
  {D.}~\bibnamefont {Ceresoli}}, \bibinfo {author} {\bibfnamefont {G.~L.}\
  \bibnamefont {Chiarotti}}, \bibinfo {author} {\bibfnamefont {M.}~\bibnamefont
  {Cococcioni}}, \bibinfo {author} {\bibfnamefont {I.}~\bibnamefont {Dabo}},
  \emph {et~al.},\ }\href@noop {} {\bibfield  {journal} {\bibinfo  {journal}
  {Journal of Physics: Condensed Matter}\ }\textbf {\bibinfo {volume} {21}},\
  \bibinfo {pages} {395502} (\bibinfo {year} {2009})}\BibitemShut {NoStop}%
\bibitem [{\citenamefont {Perdew}\ \emph {et~al.}(1996)\citenamefont {Perdew},
  \citenamefont {Burke},\ and\ \citenamefont
  {Ernzerhof}}]{perdew1996generalized}%
  \BibitemOpen
  \bibfield  {author} {\bibinfo {author} {\bibfnamefont {J.~P.}\ \bibnamefont
  {Perdew}}, \bibinfo {author} {\bibfnamefont {K.}~\bibnamefont {Burke}}, \
  and\ \bibinfo {author} {\bibfnamefont {M.}~\bibnamefont {Ernzerhof}},\
  }\href@noop {} {\bibfield  {journal} {\bibinfo  {journal} {Physical Review
  Letters}\ }\textbf {\bibinfo {volume} {77}},\ \bibinfo {pages} {3865}
  (\bibinfo {year} {1996})}\BibitemShut {NoStop}%
\bibitem [{\citenamefont {Anisimov}\ \emph {et~al.}(1991)\citenamefont
  {Anisimov}, \citenamefont {Zaanen},\ and\ \citenamefont
  {Andersen}}]{anisimov1991band}%
  \BibitemOpen
  \bibfield  {author} {\bibinfo {author} {\bibfnamefont {V.~I.}\ \bibnamefont
  {Anisimov}}, \bibinfo {author} {\bibfnamefont {J.}~\bibnamefont {Zaanen}}, \
  and\ \bibinfo {author} {\bibfnamefont {O.~K.}\ \bibnamefont {Andersen}},\
  }\href@noop {} {\bibfield  {journal} {\bibinfo  {journal} {Physical Review
  B}\ }\textbf {\bibinfo {volume} {44}},\ \bibinfo {pages} {943} (\bibinfo
  {year} {1991})}\BibitemShut {NoStop}%
\bibitem [{\citenamefont {Hsu}\ \emph {et~al.}(2009)\citenamefont {Hsu},
  \citenamefont {Umemoto}, \citenamefont {Cococcioni},\ and\ \citenamefont
  {Wentzcovitch}}]{hsu2009first}%
  \BibitemOpen
  \bibfield  {author} {\bibinfo {author} {\bibfnamefont {H.}~\bibnamefont
  {Hsu}}, \bibinfo {author} {\bibfnamefont {K.}~\bibnamefont {Umemoto}},
  \bibinfo {author} {\bibfnamefont {M.}~\bibnamefont {Cococcioni}}, \ and\
  \bibinfo {author} {\bibfnamefont {R.}~\bibnamefont {Wentzcovitch}},\
  }\href@noop {} {\bibfield  {journal} {\bibinfo  {journal} {Physical Review
  B}\ }\textbf {\bibinfo {volume} {79}},\ \bibinfo {pages} {125124} (\bibinfo
  {year} {2009})}\BibitemShut {NoStop}%
\bibitem [{\citenamefont {Timrov}\ \emph {et~al.}(2018)\citenamefont {Timrov},
  \citenamefont {Marzari},\ and\ \citenamefont
  {Cococcioni}}]{timrov2018hubbard}%
  \BibitemOpen
  \bibfield  {author} {\bibinfo {author} {\bibfnamefont {I.}~\bibnamefont
  {Timrov}}, \bibinfo {author} {\bibfnamefont {N.}~\bibnamefont {Marzari}}, \
  and\ \bibinfo {author} {\bibfnamefont {M.}~\bibnamefont {Cococcioni}},\
  }\href@noop {} {\bibfield  {journal} {\bibinfo  {journal} {Physical Review
  B}\ }\textbf {\bibinfo {volume} {98}},\ \bibinfo {pages} {085127} (\bibinfo
  {year} {2018})}\BibitemShut {NoStop}%
\bibitem [{\citenamefont {Monkhorst}\ and\ \citenamefont
  {Pack}(1976)}]{monkhorst1976special}%
  \BibitemOpen
  \bibfield  {author} {\bibinfo {author} {\bibfnamefont {H.~J.}\ \bibnamefont
  {Monkhorst}}\ and\ \bibinfo {author} {\bibfnamefont {J.~D.}\ \bibnamefont
  {Pack}},\ }\href@noop {} {\bibfield  {journal} {\bibinfo  {journal} {Physical
  Review B}\ }\textbf {\bibinfo {volume} {13}},\ \bibinfo {pages} {5188}
  (\bibinfo {year} {1976})}\BibitemShut {NoStop}%
\bibitem [{\citenamefont {Vanderbilt}(1990)}]{vanderbilt1990soft}%
  \BibitemOpen
  \bibfield  {author} {\bibinfo {author} {\bibfnamefont {D.}~\bibnamefont
  {Vanderbilt}},\ }\href@noop {} {\bibfield  {journal} {\bibinfo  {journal}
  {Physical Review B}\ }\textbf {\bibinfo {volume} {41}},\ \bibinfo {pages}
  {7892} (\bibinfo {year} {1990})}\BibitemShut {NoStop}%
\bibitem [{\citenamefont {Henkelman}\ \emph {et~al.}(2000)\citenamefont
  {Henkelman}, \citenamefont {Uberuaga},\ and\ \citenamefont
  {J{\'o}nsson}}]{henkelman2000climbing}%
  \BibitemOpen
  \bibfield  {author} {\bibinfo {author} {\bibfnamefont {G.}~\bibnamefont
  {Henkelman}}, \bibinfo {author} {\bibfnamefont {B.~P.}\ \bibnamefont
  {Uberuaga}}, \ and\ \bibinfo {author} {\bibfnamefont {H.}~\bibnamefont
  {J{\'o}nsson}},\ }\href@noop {} {\bibfield  {journal} {\bibinfo  {journal}
  {The Journal of Chemical Physics}\ }\textbf {\bibinfo {volume} {113}},\
  \bibinfo {pages} {9901} (\bibinfo {year} {2000})}\BibitemShut {NoStop}%
\bibitem [{\citenamefont {King-Smith}\ and\ \citenamefont
  {Vanderbilt}(1993)}]{king1993theory}%
  \BibitemOpen
  \bibfield  {author} {\bibinfo {author} {\bibfnamefont {R.}~\bibnamefont
  {King-Smith}}\ and\ \bibinfo {author} {\bibfnamefont {D.}~\bibnamefont
  {Vanderbilt}},\ }\href@noop {} {\bibfield  {journal} {\bibinfo  {journal}
  {Physical Review B}\ }\textbf {\bibinfo {volume} {47}},\ \bibinfo {pages}
  {1651} (\bibinfo {year} {1993})}\BibitemShut {NoStop}%
\bibitem [{\citenamefont {Resta}(1993)}]{resta1993macroscopic}%
  \BibitemOpen
  \bibfield  {author} {\bibinfo {author} {\bibfnamefont {R.}~\bibnamefont
  {Resta}},\ }\href@noop {} {\bibfield  {journal} {\bibinfo  {journal}
  {Europhysics Letters}\ }\textbf {\bibinfo {volume} {22}},\ \bibinfo {pages}
  {133} (\bibinfo {year} {1993})}\BibitemShut {NoStop}%
\bibitem [{\citenamefont {Vanderbilt}(1993)}]{vanderbilt1993d}%
  \BibitemOpen
  \bibfield  {author} {\bibinfo {author} {\bibfnamefont {D.}~\bibnamefont
  {Vanderbilt}},\ }\href@noop {} {\bibfield  {journal} {\bibinfo  {journal}
  {Physycal Review B}\ }\textbf {\bibinfo {volume} {48}},\ \bibinfo {pages}
  {4442} (\bibinfo {year} {1993})}\BibitemShut {NoStop}%
\bibitem [{\citenamefont {Dronskowski}\ and\ \citenamefont
  {Bl{\"o}chl}(1993)}]{dronskowski1993crystal}%
  \BibitemOpen
  \bibfield  {author} {\bibinfo {author} {\bibfnamefont {R.}~\bibnamefont
  {Dronskowski}}\ and\ \bibinfo {author} {\bibfnamefont {P.~E.}\ \bibnamefont
  {Bl{\"o}chl}},\ }\href@noop {} {\bibfield  {journal} {\bibinfo  {journal}
  {The Journal of Physical Chemistry}\ }\textbf {\bibinfo {volume} {97}},\
  \bibinfo {pages} {8617} (\bibinfo {year} {1993})}\BibitemShut {NoStop}%
\bibitem [{\citenamefont {Deringer}\ \emph {et~al.}(2011)\citenamefont
  {Deringer}, \citenamefont {Tchougr{\'e}eff},\ and\ \citenamefont
  {Dronskowski}}]{deringer2011crystal}%
  \BibitemOpen
  \bibfield  {author} {\bibinfo {author} {\bibfnamefont {V.~L.}\ \bibnamefont
  {Deringer}}, \bibinfo {author} {\bibfnamefont {A.~L.}\ \bibnamefont
  {Tchougr{\'e}eff}}, \ and\ \bibinfo {author} {\bibfnamefont {R.}~\bibnamefont
  {Dronskowski}},\ }\href@noop {} {\bibfield  {journal} {\bibinfo  {journal}
  {The Journal of Physical Chemistry A}\ }\textbf {\bibinfo {volume} {115}},\
  \bibinfo {pages} {5461} (\bibinfo {year} {2011})}\BibitemShut {NoStop}%
\bibitem [{\citenamefont {Maintz}\ \emph {et~al.}(2013)\citenamefont {Maintz},
  \citenamefont {Deringer}, \citenamefont {Tchougr{\'e}eff},\ and\
  \citenamefont {Dronskowski}}]{maintz2013analytic}%
  \BibitemOpen
  \bibfield  {author} {\bibinfo {author} {\bibfnamefont {S.}~\bibnamefont
  {Maintz}}, \bibinfo {author} {\bibfnamefont {V.~L.}\ \bibnamefont
  {Deringer}}, \bibinfo {author} {\bibfnamefont {A.~L.}\ \bibnamefont
  {Tchougr{\'e}eff}}, \ and\ \bibinfo {author} {\bibfnamefont {R.}~\bibnamefont
  {Dronskowski}},\ }\href@noop {} {\bibfield  {journal} {\bibinfo  {journal}
  {Journal of Computational Chemistry}\ }\textbf {\bibinfo {volume} {34}},\
  \bibinfo {pages} {2557} (\bibinfo {year} {2013})}\BibitemShut {NoStop}%
\bibitem [{\citenamefont {Maintz}\ \emph
  {et~al.}(2016{\natexlab{a}})\citenamefont {Maintz}, \citenamefont {Esser},\
  and\ \citenamefont {Dronskowski}}]{maintz2016efficient}%
  \BibitemOpen
  \bibfield  {author} {\bibinfo {author} {\bibfnamefont {S.}~\bibnamefont
  {Maintz}}, \bibinfo {author} {\bibfnamefont {M.}~\bibnamefont {Esser}}, \
  and\ \bibinfo {author} {\bibfnamefont {R.}~\bibnamefont {Dronskowski}},\
  }\href@noop {} {\bibfield  {journal} {\bibinfo  {journal} {Acta Physica
  Polonica B}\ }\textbf {\bibinfo {volume} {47}} (\bibinfo {year}
  {2016}{\natexlab{a}})}\BibitemShut {NoStop}%
\bibitem [{\citenamefont {Maintz}\ \emph
  {et~al.}(2016{\natexlab{b}})\citenamefont {Maintz}, \citenamefont {Deringer},
  \citenamefont {Tchougr{\'e}eff},\ and\ \citenamefont
  {Dronskowski}}]{maintz2016lobster}%
  \BibitemOpen
  \bibfield  {author} {\bibinfo {author} {\bibfnamefont {S.}~\bibnamefont
  {Maintz}}, \bibinfo {author} {\bibfnamefont {V.~L.}\ \bibnamefont
  {Deringer}}, \bibinfo {author} {\bibfnamefont {A.~L.}\ \bibnamefont
  {Tchougr{\'e}eff}}, \ and\ \bibinfo {author} {\bibfnamefont {R.}~\bibnamefont
  {Dronskowski}},\ }\href@noop {} {\bibfield  {journal} {\bibinfo  {journal}
  {Journal of Computational Chemistry}\ }\textbf {\bibinfo {volume} {37}},\
  \bibinfo {pages} {1030} (\bibinfo {year} {2016}{\natexlab{b}})}\BibitemShut
  {NoStop}%
\bibitem [{\citenamefont {Momma}\ and\ \citenamefont
  {Izumi}(2011)}]{momma2011vesta}%
  \BibitemOpen
  \bibfield  {author} {\bibinfo {author} {\bibfnamefont {K.}~\bibnamefont
  {Momma}}\ and\ \bibinfo {author} {\bibfnamefont {F.}~\bibnamefont {Izumi}},\
  }\href@noop {} {\bibfield  {journal} {\bibinfo  {journal} {Journal of Applied
  Crystallography}\ }\textbf {\bibinfo {volume} {44}},\ \bibinfo {pages} {1272}
  (\bibinfo {year} {2011})}\BibitemShut {NoStop}%
\bibitem [{\citenamefont {Goldschmidt}(1926)}]{goldschmidt1926gesetze}%
  \BibitemOpen
  \bibfield  {author} {\bibinfo {author} {\bibfnamefont {V.~M.}\ \bibnamefont
  {Goldschmidt}},\ }\href@noop {} {\bibfield  {journal} {\bibinfo  {journal}
  {Naturwissenschaften}\ }\textbf {\bibinfo {volume} {14}},\ \bibinfo {pages}
  {477} (\bibinfo {year} {1926})}\BibitemShut {NoStop}%
\bibitem [{\citenamefont {Benedek}\ and\ \citenamefont
  {Fennie}(2013)}]{benedek2013there}%
  \BibitemOpen
  \bibfield  {author} {\bibinfo {author} {\bibfnamefont {N.~A.}\ \bibnamefont
  {Benedek}}\ and\ \bibinfo {author} {\bibfnamefont {C.~J.}\ \bibnamefont
  {Fennie}},\ }\href@noop {} {\bibfield  {journal} {\bibinfo  {journal} {The
  Journal of Physical Chemistry C}\ }\textbf {\bibinfo {volume} {117}},\
  \bibinfo {pages} {13339} (\bibinfo {year} {2013})}\BibitemShut {NoStop}%
\bibitem [{\citenamefont {Aschauer}\ and\ \citenamefont
  {Spaldin}(2014)}]{aschauer2014competition}%
  \BibitemOpen
  \bibfield  {author} {\bibinfo {author} {\bibfnamefont {U.}~\bibnamefont
  {Aschauer}}\ and\ \bibinfo {author} {\bibfnamefont {N.~A.}\ \bibnamefont
  {Spaldin}},\ }\href@noop {} {\bibfield  {journal} {\bibinfo  {journal}
  {Journal of Physics: Condensed Matter}\ }\textbf {\bibinfo {volume} {26}},\
  \bibinfo {pages} {122203} (\bibinfo {year} {2014})}\BibitemShut {NoStop}%
\bibitem [{\citenamefont {Attfield}(2013)}]{attfield2013principles}%
  \BibitemOpen
  \bibfield  {author} {\bibinfo {author} {\bibfnamefont {J.~P.}\ \bibnamefont
  {Attfield}},\ }\href@noop {} {\bibfield  {journal} {\bibinfo  {journal}
  {Crystal Growth \& Design}\ }\textbf {\bibinfo {volume} {13}},\ \bibinfo
  {pages} {4623} (\bibinfo {year} {2013})}\BibitemShut {NoStop}%
\bibitem [{\citenamefont {Denis~Romero}\ \emph {et~al.}(2014)\citenamefont
  {Denis~Romero}, \citenamefont {Leach}, \citenamefont {M{\"o}ller},
  \citenamefont {Foronda}, \citenamefont {Blundell},\ and\ \citenamefont
  {Hayward}}]{denis2014strontium}%
  \BibitemOpen
  \bibfield  {author} {\bibinfo {author} {\bibfnamefont {F.}~\bibnamefont
  {Denis~Romero}}, \bibinfo {author} {\bibfnamefont {A.}~\bibnamefont {Leach}},
  \bibinfo {author} {\bibfnamefont {J.~S.}\ \bibnamefont {M{\"o}ller}},
  \bibinfo {author} {\bibfnamefont {F.}~\bibnamefont {Foronda}}, \bibinfo
  {author} {\bibfnamefont {S.~J.}\ \bibnamefont {Blundell}}, \ and\ \bibinfo
  {author} {\bibfnamefont {M.~A.}\ \bibnamefont {Hayward}},\ }\href@noop {}
  {\bibfield  {journal} {\bibinfo  {journal} {Angewandte Chemie International
  Edition}\ }\textbf {\bibinfo {volume} {53}},\ \bibinfo {pages} {7556}
  (\bibinfo {year} {2014})}\BibitemShut {NoStop}%
\bibitem [{\citenamefont {Huster}(1980)}]{huster1980kristallstruktur}%
  \BibitemOpen
  \bibfield  {author} {\bibinfo {author} {\bibfnamefont {J.}~\bibnamefont
  {Huster}},\ }\href@noop {} {\bibfield  {journal} {\bibinfo  {journal}
  {Zeitschrift f{\"u}r Naturforschung B}\ }\textbf {\bibinfo {volume} {35}},\
  \bibinfo {pages} {775} (\bibinfo {year} {1980})}\BibitemShut {NoStop}%
\bibitem [{\citenamefont {Beckman}\ \emph {et~al.}(2009)\citenamefont
  {Beckman}, \citenamefont {Wang}, \citenamefont {Rabe},\ and\ \citenamefont
  {Vanderbilt}}]{beckman2009ideal}%
  \BibitemOpen
  \bibfield  {author} {\bibinfo {author} {\bibfnamefont {S.}~\bibnamefont
  {Beckman}}, \bibinfo {author} {\bibfnamefont {X.}~\bibnamefont {Wang}},
  \bibinfo {author} {\bibfnamefont {K.~M.}\ \bibnamefont {Rabe}}, \ and\
  \bibinfo {author} {\bibfnamefont {D.}~\bibnamefont {Vanderbilt}},\
  }\href@noop {} {\bibfield  {journal} {\bibinfo  {journal} {Physical Review
  B}\ }\textbf {\bibinfo {volume} {79}},\ \bibinfo {pages} {144124} (\bibinfo
  {year} {2009})}\BibitemShut {NoStop}%
\end{thebibliography}%

\end{document}